\documentclass[aps,prl,twocolumn,preprintnumbers,10pt,showpacs]{revtex4-1}
\usepackage{epsfig,psfrag,amsmath,amssymb,lmodern}
\usepackage[normalem]{ulem}
\usepackage{color}
\usepackage[applemac]{inputenc}

\usepackage[T2A,T1]{fontenc}

\allowdisplaybreaks[4]

\def \nn {\nonumber}

\def \e  {\mathop{\rm e}\nolimits}

\newcommand \ket [1] {|{#1}\rangle}
\newcommand \bra [1] {\langle {#1}|}

\newcommand{\cN}{{\cal N}}

\newcommand{\cO}{{\cal O}}

\newcommand{\z}{\zeta}

\newcommand \vev [1] {\langle{#1}\rangle}

\newcommand{\ab}[1]{\langle #1 \rangle}

\definecolor{darkred}{rgb}{0.5,0.0,0.0}
\definecolor{darkblue}{rgb}{0.0,0.0,0.9}
\definecolor{darkerblue}{rgb}{0.0,0.0,0.5}
\definecolor{darkgreen}{rgb}{0.0,0.5,0.0}
\definecolor{black}{rgb}{0.0,0.0,0.0}
\definecolor{brown}{rgb}{0.6,0.4,0.2}
\newcommand{\red}{\color{darkred}}
\newcommand{\blue}{\color{darkblue}}

\def \be  {\begin{equation}}
\def \ee  {\end{equation}}
\def \ba  {\begin{eqnarray}}
\def \ea  {\end{eqnarray}}
\def \baa {\begin{eqnarray*}}
\def \eaa {\end{eqnarray*}}
\def \bb  {\begin {thebibliography} }
\def \eb  {\end{thebibliography}}

\makeatletter
\renewcommand*\env@matrix[1][*\c@MaxMatrixCols c]{%
  \hskip -\arraycolsep
  \let\@ifnextchar\new@ifnextchar
  \array{#1}}
\makeatother

\begin{document}
\title{
Three-point  energy correlator in $\mathcal{N}=4$ super-Yang Mills Theory
}

\author{ 
Kai  Yan$^{a}$,  Xiaoyuan Zhang$^{b}$
}

\affiliation{
$^a$ Shanghai Jiao Tong University, 800 Dongchuan Road, 200240 Shanghai, China\\ 
$^b$ Harvard University, 02138 Cambridge, MA, US 
}

\begin{abstract}
An analytic formula is given for the three-point energy correlator (EEEC) at leading order (LO) in maximally supersymmetric Yang-Mills theory ($\mathcal{N}=4$ sYM). This is the first analytic calculation of a three-parameter event shape observable, which provides valuable data for various studies ranging from conformal field theories to jet substructure. The associated class of functions  define a new type of single-valued polylogarithms characterized by 16 alphabet letters, which 
 manifest a $D_6 \times Z_2$ dihedral symmetry of the event shape. 
With the unexplored simplicity in the perturbative structure of EEEC, all kinematic regions including collinear, squeezed and coplanar limits are now available.

\end{abstract}

\maketitle


\noindent \textbf{1. Introduction. }  

The energy correlator observable measures the energy deposited in multiple detectors as a function of angles between the detectors. From the phenomenological perspective, energy correlators probe the energy flow and can be used as jet observables \cite{Coleman:2017fiq,Larkoski:2017jix,Asquith:2018igt,Komiske:2018cqr,Marzani:2019hun,Datta:2019ndh} for precise tests of the standard model or new physics search. From the practical side, energy correlators is perhaps the simplest infared safe event shape \cite{Kinoshita:1962ur,Lee:1964is} to calculate analytically. From the theory side, they belong to class of observables probing the spatial correlation among flow operators, which provides valuable data for understanding the nature of quantum field theories \cite{Hofman:2008ar,Arkani-Hamed:2015bza,Korchemsky:2021okt}. 

 The two-point energy correlator (EEC) \cite{Basham:1978bw} is computed analytically to next-to-leading order (NLO) in quantum chromodynamics (QCD) \cite{Dixon:2018qgp, Luo:2019nig} and NNLO in $\cN=4$ super-Yang-Mills theory (sYM) \cite{Belitsky:2013ofa,Henn:2019gkr},  numerically up to NNLO in QCD  \cite{Richards:1982te,Richards:1983sr,Ali:1982ub,Falck:1988gb,Kunszt:1989km,Glover:1994vz,Clay:1995sd,Kramer:1996qr,DelDuca:2016csb,DelDuca:2016ily}, and resummed to all orders  in both the back-to-back  \cite{Collins:1981uk,Kodaira:1981nh,deFlorian:2004mp,Tulipant:2017ybb,Moult:2018jzp, Moult:2019vou} and collinear limit \cite{Konishi:1978yx,Konishi:1978ax, Dixon:2019uzg}. Meanwhile the precision study on multi-particle energy correlator has been initiated, featuring
  the LO prediction for the three-point energy correlator (EEEC) in the triple-collinear limit \cite{Chen:2019bpb}.

The EEEC, which depends on three angles among the detectors, captures the nontrivial shape dependence in the scattering processes.
The standard definition for the EEEC as a differential cross-section can then be recast as a five-point 
 correlation function 
\begin{align}\label{def-EEC-main}
{\rm EEEC} (\chi_1, \chi_2,\chi_3) =&  \int \prod_{i=1}^3  \left[ d \Omega_{\vec n_i} \delta(\vec n_i \cdot \vec n_{i+1} - \cos \chi_i ) \right]  \nonumber \\
& \hspace{-2.5cm} \times
\frac{\int d^4 x \e^{iqx} \vev{0|O^\dagger (x){\cal E}(\vec n_1){\cal E}(\vec n_2) {\cal E}(\vec n_3) O(0) |0}} { (q^0)^3 \int d^4 x \e^{iqx} \vev{0|O^\dagger (x) O(0) |0}} \,.
\end{align}
Here the detector operator that measures the energy flux in the direction $\vec n$ is given by
an integrated stress-energy tensor $T_{\mu\nu}$  \cite{Sveshnikov:1995vi,Korchemsky:1997sy,Korchemsky:1999kt,Belitsky:2001ij,Hofman:2008ar}, ${\cal E}(\vec n)  = \int_{- \infty}^{\infty} d \tau \ \lim_{r \to \infty} r^2  n^i T_{0i}(t=\tau+r, r \vec n) \, .$
The operators $O$ (source) and $O^\dagger$ (sink) create the final state,  
whose particles are detected by the two calorimeters. 
The choice of the local operator $O$ depends on the physical problem. 
For  $e^+ e^-$ annihilation, $O$ is given by an electromagnetic current.

\begin{figure}[t]
\includegraphics[width = 0.25\textwidth]{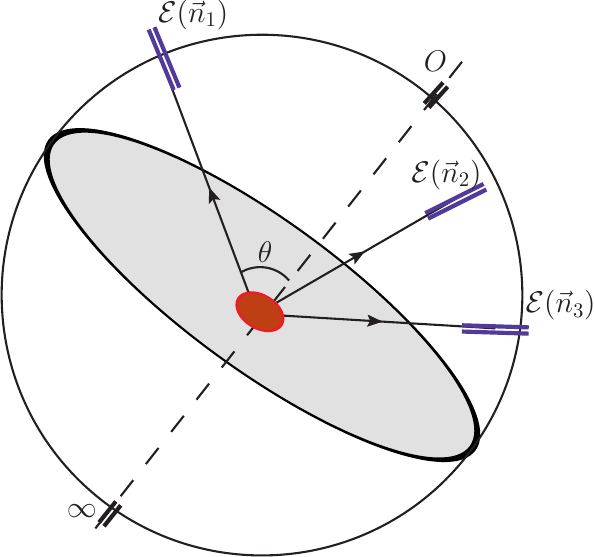} \quad
\includegraphics[width = 0.2\textwidth]{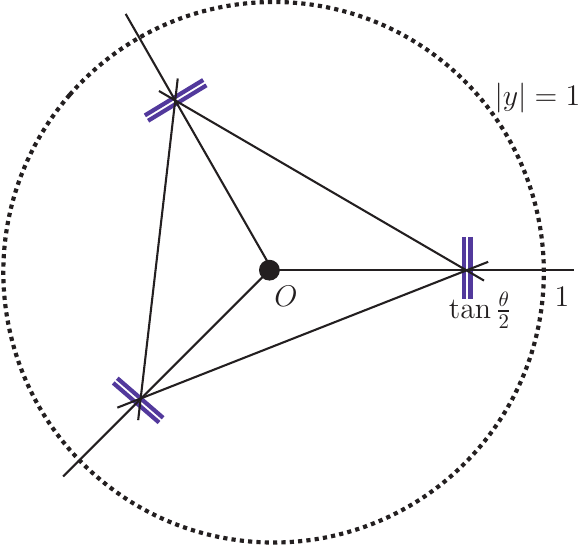}
\caption{\small Graphical representation of the three-point energy correlator: particles produced out of the vacuum by
the source are captured by the three detectors located at spatial infinity in the directions of the unit vectors $\vec n_1, \vec n_2 $ and $\vec n_3$. They can be mapped onto three points located on a circle with radius $|y|=\tan \frac{\theta}{2}$ on the celestial sphere. 
The three angles are parametrized  by 
 $(\sin \frac{\chi_1}{2},\sin \frac{\chi_2}{2}, \sin \frac{\chi_3}{2}) = \sin \theta \,( \sin\frac{\phi_1}{2}, \sin \frac{\phi_2}{2}, \sin \frac{\phi_1+\phi_2}{2})$
.}
\label{fig:EEEC}
\end{figure}

The energy correlator is an onshell observable that bares close relations to  the offshell correlation functions involving the stress-energy tensors. 
In view of this property, previous studies  in  $\mathcal{N}=4$ sYM theory employed various shortcut to obtaining the EEC by taking multiple discontinuities of  Euclidean correlation functions and by exploiting the superconformal symmetries of the latter \cite{Belitsky:2013ofa,Henn:2019gkr,Belitsky:2013bja,Belitsky:2014zha,Korchemsky:2015ssa}
.
It remains unknown whether such approaches are feasible for computing the higher-point energy correlators.   
In this letter we adopt an onshell approach to obtain the EEEC from the super form factor for protected scalar operator \cite{Eden:2011yp,
Eden:2011ku},  benefiting from the simplicity of matrix elements  in $\mathcal{N}=4$ sYM.   
We present the one-loop EEEC result for arbitrary angles, which is the first analytic calculation of multi-particle correlation observables with full shape dependence. 

Our result encodes valuable information on the function space of the EEEC in perturbative quantum field theory: the classifications of symmetries, symbol alphabets and polylogarithms. These mathematical structures are studied much more thoroughly in the context of scattering amplitudes than finite observables in collider experiments.   
We are strongly motivated to initiate the discussion on these topics for the energy correlator observable, starting from $\mathcal{N}=4$ sYM.
As 
they provide powerful tools and experience for QCD \cite{Henn:2020omi}, which is phenomenologically relevant for the cutting edge studies at LHC.

  \medskip
 \noindent \textbf{2. EEEC from four point form factor.}


In $\cN=4$ sYM, we may choose the source and sink 
to be scalar operators that are the bottom component of the supermultiplet of conserved currents.
As such, they are natural analogs of the electromagnetic current, and have fixed conformal weight two. 
The matrix element for producing  a given onshell super-state from the vacumm defines the so-called form factor \cite{vanNeerven:1985ja,
Bork:2010wf,Bork:2011cj,
Penante:2014sza,Brandhuber:2011tv}
\begin{align}
 \int d^4 x  \, e^{i q \cdot x } \,  \bra{X} O(x) \ket{0}  \equiv (2 \pi)^4 \delta^{4}(q- p_X) \, F_{X} 
\end{align} 

 In perturbative theories EEEC  
 can be obtained from the squared form factor by performing a weighted sum over the  onshell external states. 
For convenience  we normalize the event shape by a volume factor of the phase space, 
thus defining a function $H$ through
  $\text{ EEEC} (\chi_1, \chi_2, \chi_3) \equiv  (8 \pi^2 )  \times  || \vec{n}_1 \wedge \vec n_2  \wedge \vec{n}_3 ||^{-1}\,  H(\vec{n}_1, \vec{n}_2 , \vec{n}_3 ) $,  which we can evaluate by carrying out the onshell phase-space integration while fixing the directions of three particles in the final states
 \begin{align}
 H( \vec{n}_1, \vec{n}_2, \vec{n}_3) & = \frac{1}{\sigma_{\rm tot}}
\sum_{(i,j,k) \in X} \int d \Pi_{X}  \delta^2 ( \vec{n}_1 - \hat{p}_i) \,
 \\
& \times  \delta^2 ( \vec{n}_2 - \hat{p}_j) \,
  \delta^2 ( \vec{n}_3 - \hat{p}_k)  \, \frac{E_i E_j E_k}{(q^0)^3} \, |F_X|^2   \nn 
 \end{align}
where $i, j$ and $k$ run over all final-state particles.
$H$ has the perturbative expansion
$H = \sum_{k \ge 1} a^k H^{(k)}$
 in the  `t Hooft coupling. 
The born level 
event shape is a delta function due to 3-body kinematic constraints, 
$H_{\rm Born} = 
  \delta (|| \vec{n}_1 \wedge \vec n_2  \wedge \vec{n}_3 ||)  \, \csc^2 \frac{\chi_1}{2} \csc^2 \frac{\chi_2}{2} \csc^2 \frac{\chi_3}{2}\,.
$ 
The leading order that has nontrivial three-angle dependence  is $\cO(a^2)$, where the complication comes from the tree-level
 four-point NMHV super matrix elements $|F_4|^2$, for details see \cite{Bianchi:2018peu,FF1}.  
 After summing over super states and symmetrization over the final-state momenta, the squared four-point form factor can be organized into a concise form involving dual conformal cross ratios, 
  \begin{align}
 |F_4|_{\rm sym.}^2 &= \frac{q^4}{ s_{12} s_{23} s_{34}s_{41}} \left[ \frac14 +
 2 \frac{ s_{23} s_{41} }{ s_{341} s_{412} } +  
  \frac{s_{23} s_{34}}{  s_{123}s_{341}  }  \right] \nonumber \\ & + \text{perm} (1,2,3,4) .
 \end{align} 
 where the second and third term in the bracket correspond to the NMHV contribution. 

To compute the EEEC, we first apply topology identification to the squared matrix elements. With the EEEC measurement functions, they can be decomposed into four topologies. 
To proceed, we parameterize the kinematic invariants by the energy fractions of three detected final-state particles as well as the three angles $(\theta, \phi_1, \phi_2)$ as depicted in Fig.~\ref{fig:EEEC}.
In particular, we switch to the following set of angle parameters 
\begin{align} \label{anglecoords}
s = \tan^{2} \frac{\theta}{2}, \quad \tau_1 = e^{i \phi_1}, \quad \tau_2 = e^{i \phi_2} 
\end{align} 
such that both the matrix elements and phase space simplify down. 
Integrating the four-particle phase space \cite{Gehrmann-DeRidder:2003pne} against the measurement functions, we are left with a set of two-fold integrals which are linearly reducible \cite{Brown:2008um,Brown:2009ta,Bogner:2013tia,Panzer:2015ida},which allow us to
compute directly in \texttt{HyperInt} \cite{Panzer:2014caa}.

 

\begin{figure}[t]
\includegraphics[width=0.21\textwidth]{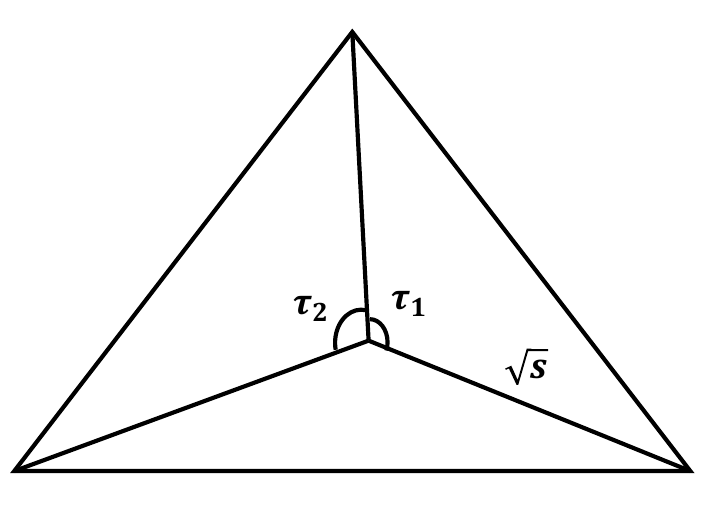}
\includegraphics[width = 0.26\textwidth]{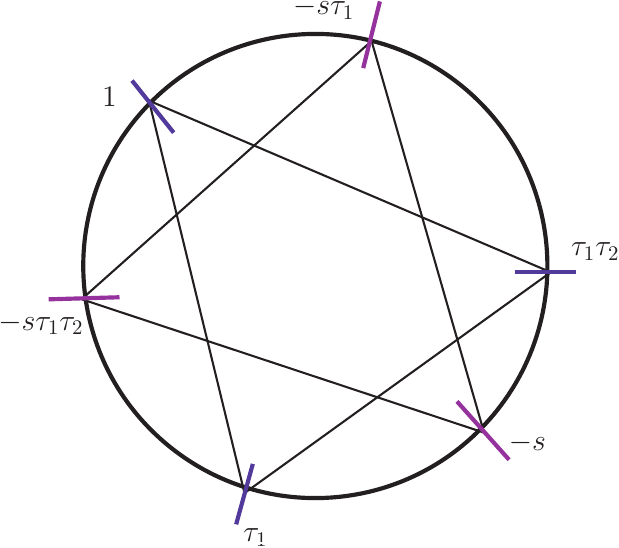}
\caption{\small  Embedding of the EEEC kinematics.   Left: (a) three points on the celestial sphere which we put on a unit circle with radius $\sqrt{s}$ centered at the origin. 
 Right: (b) the realization of this kinematic configuration as a hexagon located on a unit circle. 
}
\label{fig:eee-polygons}
\end{figure}

 \medskip 
 \noindent \textbf{3. Symbol alphabets}
 
The EEEC can be expressed in a frame independent manner as a function of three conformally invariant variables 
\begin{align}\label{zijdef}
\zeta_{ij} = \frac{ q^2(p_i \cdot p_j)}{ 2(q\cdot p_i ) (q \cdot p_j)} = \frac{ \ab{p_i p_j} \ab{\xi_j \xi_i} }{ \ab{p_i \xi_i} \ab{p_j \xi_j}}\, , \;\; 
\ket{\xi_j}  \equiv  q |j] 
\end{align}
From the the results of function $H$  we read off 16 symbol alphabets,  which contains two types of algebraic roots 
\begin{align}
|\Delta_1| &\equiv  \big{|}  \big{|} \vec{n}_1 \wedge \vec{n}_2 \wedge \vec{n}_3  \big{|}  \big{|}  =| \sqrt{(1- u_1-u_2-u_3)^2 - 4 u_1u_2 u_3}|
\nn \\
 |\Delta_2 |& \equiv  \big{|}  \big{|}  \vec{n}_1  \wedge \vec{n}_2  +  \vec{n}_2  \wedge \vec{n}_3+ \vec{n}_3  \wedge \vec{n}_1 \big{|}  \big{|}   =
   \big| \sqrt{\lambda (\zeta_{12} , \zeta_{23}, \zeta_{31})  } \, \big| \nn
\end{align}
where $\{ u_i \}  \equiv \{ 1- \zeta_{ij} \} $, and $\lambda(a,b,c) \equiv a^2+b^2+c^2-2 ab- 2 ac-2bc $ is the K{\'a}llen function. 

The representation of variables in Eq.~\eqref{zijdef} foreshadows a $D_6$ dihedral symmmetry of the EEEC, which can be better visualized as we 
 embed the kinematic data $ \ket{p_i} \equiv \ket{2i -1},  \ket{\xi_i}  \equiv \ket{2i+2} $ in a $2 \times 6$ matrix $Z$:   $Z_a  = \ket{a} \in  CP^1 $, 
\begin{align}
Z  \doteq  \begin{pmatrix}
 1 &    1    &  1 & 1     & 1                      & 1                                \\ 
 \eta_1 &  \eta_2   & \eta_3  &  \eta_4   &  \eta_5  & \eta_6
\end{pmatrix}   , \;  \eta_{a}- \eta_b \doteq \frac{ \vev{ab}}{\vev{a \infty} \vev{\infty b}}
\end{align}
where $ \ket{\infty} = (0,1)^{T}$.   

The geometric interpretation of the above formalism is clear 
in the center of mass frame where $q= (q^0,0,0,0)$, so that under 
steoreographic projection the three vectors $\vec{p}_i$  
are mapped onto a triangle $(y_i, \bar y_i) = (\eta_i, - 1/ \eta_{i+3}) $ on the celestial sphere. 
Let us further introduce a special point  $(y_I, \bar{y}_{I})= (\eta_I, -1/\eta_{\bar I})$  representing  the center 
of the triangle, 
whose location determined by the equations 
\begin{align}
\frac{ \ab{1 I} \ab{4 \bar I } \ab{52}}{ \ab{14} \ab{ 2 \bar I } \ab{5 I} } = 
\frac{ \ab{3 I} \ab{6 \bar I } \ab{14}}{ \ab{36} \ab{1 I } \ab{4 \bar I} } =1
\label{eq:circumcenter}
\end{align} 
We may impose $ \ket{I} = (0,1)^T,  \ket{\bar I} = (1,0)^T$ and  $\vev{ \bar I 1}= \vev{aI} = 1$, and expand the kinematic space to include $I$ (or equivalently $\bar I$).  Thus  we fix the gauge under which  
\begin{align}
\hskip-0.2mm
 Z \Big| I = \begin{pmatrix}[cccccc|c]
 1 &    1    &  1 & 1     & 1                      & 1                              & 0  \\ 
 1 &  -s \tau_1 \tau_2   &   \tau_1 &  -s  &   \tau_1 \tau_2   &  -s \tau_1  &1
\end{pmatrix} 
\label{eq:embed}
\end{align}
The kinematic space is mapped onto two similar  triangles whose circumcenter sit at the origin, which we display as a hexagon located on a unit circle FIG.~\ref{fig:eee-polygons}b.  

In this way we identify all 16 EEEC alphabets with products of pl\"{u}cker variables  $\vev{ab}, \vev{aI}$  
 as well as certain homogeneous polynomials in the form  
\begin{align}
 d_{(ab) (cd) (ef)} \equiv  \ab{a d} \ab{e b} \ab{c f} -    \ab{a f} \ab{c b} \ab{e d}
\end{align} 
Switching variables into 3 conformally invariant  ratios,  
\begin{align} 
 & y = - \frac{\ab{31} \ab{5  I} }{\ab{15}  \ab{I3} }, \quad  z = - \frac{\ab{13} \ab{56} }{ \ab{35} \ab{61} },\quad 
  w = \frac{\ab{51} \ab{62} \ab{43} }{ \ab{35} \ab{16} \ab{24} } \,.  \nn
\end{align}
we could  transform the alphabets  into 16  polynomials with all positive signs, which read
\begin{align} \label{eeeltrs}
 \Big\{ &  w \, ,  1 + w \,,  y \,,  1 + y \, , z\, , 1 + z \, ,  w + z\, , 1 + w + z \,,  \nn \\
 & y + z + y z\, ,   w +y+ z + y z\, , 1 + w + z + y z\, , \nn \\
 & 1 + w +y+ 2 z + y z\, ,  y + w y +y^2 +z+ 2 y z + y^2 z\, , \nn\\
 & 1 +y   + w y +y^2 +z+ 2 y z + y^2 z\, , \nn \\
 &1 + w +y+ w y +y^2+ z+ 2 y z + y^2 z\, ,  \nn \\
 &1 + w +y+ w y +y^2 +2 z+  2 y z + y^2 z  \, \Big\} 
\end{align}

\begin{figure}[t] 
\includegraphics[scale=0.28]{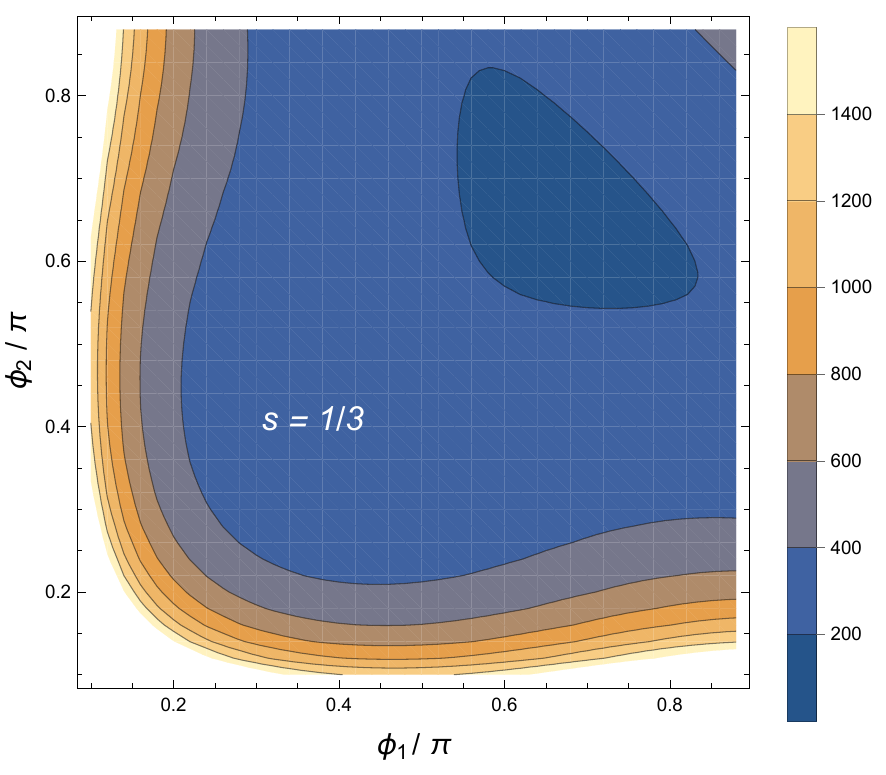}
\includegraphics[scale=0.28]{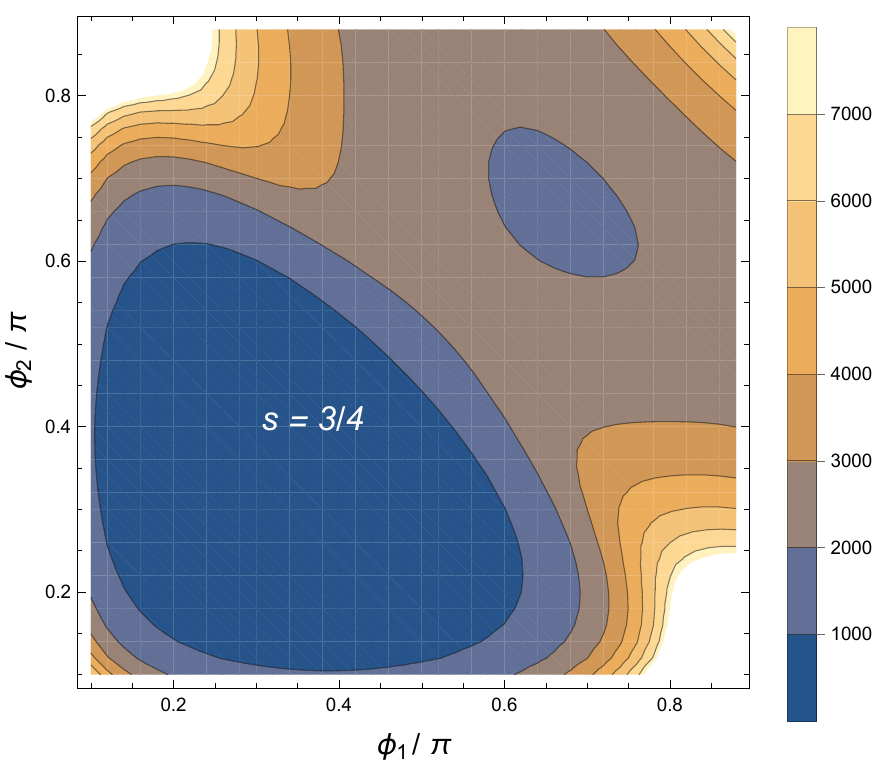} 
\includegraphics[width=0.35\textwidth]{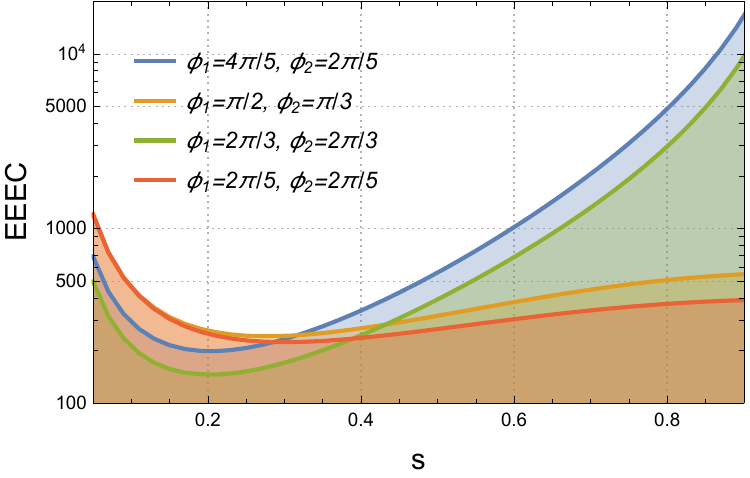} 
\caption{\small   EEEC at LO in $\mathcal{N}=4$ sYM. 
We display the function $H (s, \tau_1 = e^{i \phi_1}, \tau_2= e^{i \phi_2})$ in various kinematic regions. Top: distribution in $(\phi_1, \phi_2)$ with $s=$constant. Bottom: distribution in $s$ with fixed $(\phi_1, \phi_2)$.   } \label{fig2}
\end{figure}

%
%
%

%

 \medskip 
   \noindent \textbf{4. Symmetries and functional basis.}

Within our embedding formalism, 
  the EEEC exhibits a set of  discrete symmetries. 
 First we have   $D_6$ dihedral symmetries  acting on the hexagon coordinates $Z_a \,(a+6 \doteq a)$, which are generated by dihedral flip $\tau$: $  a \xrightarrow{\tau} 4- a $;  cyclic permutation $\sigma$:   $a \xrightarrow{\sigma} a+2$; as well as parity conjugation $P$: $ a \xrightarrow{P}  a+3 $. 
In addition, there is a residual $Z_2$ symmetry correponding  to the exchange between two solutions to Eq.~\eqref{eq:circumcenter}. 
It is generated by an operation which we call reflection $R:  \frac{\ab{I a}}{ \ab{I a+2} }  \xrightarrow{R} \frac{\ab{a\, a+3}   \ab{I a+5 }  }{ \ab{a+2 \, a+5} \ab{I a+3 }  }$.  
 $P$ and $R$ flip the signs of $\Delta_1$ and $\Delta_2$, respectively, 
 such that $\Delta_1 \xrightarrow{P} - \Delta_1, \Delta_1  \xrightarrow{R} \Delta_1$, while $ \Delta_2 \xrightarrow{P} \Delta_2, \Delta_2 \xrightarrow{R}  -\Delta_2$. 


In light of these properties, we are ready to lift the one-loop symbols into polylogarithmic functions. 
To start, we shall identify a set of variables $S$ which is closed under $\{ \tau, \sigma, P , R \}$,  such that 
 $\{ S, 1+S\}$ factorize into polynomials that cover the EEEC 
  alphabet letters in Eq.~\eqref{eeeltrs}. 

Let us introduce these variables. 
For clarity, we switch to a different gauge by performing a $GL(2)\times GL(1)^6$  transformation  on  Eq.~\eqref{eq:embed}, 
\begin{align}
Z\Big|I  =    
\begin{pmatrix}[cccccc|c]
 1 & 1 &  1 & 0        &   1             & 1                            &     1 \\ 
 0 & \frac{1}{1+x_1} &  1& - 1 &  -x_2 &  -\frac{x_2 x_3}{ 1+x_3 } & \frac{1+x_4}{x_4}
\end{pmatrix}    
\end{align}
thus introducing four parameters $(x_1, x_2, x_3, x_4)$,  three of which being independent.  

Notice that the hexagon $Z$ corresponds to the grassmaniann $\rm{Gr}(2, 6)/[GL(1)]^5$, 
which can be associated with a $A_3-$cluster algebra, with a quiver being $(x_1, x_2, x_3)$.  
Given this observation,
we introduce 15 conformally invariant ratios to cover the full set of  $\mathcal{X}-$coordinates \cite{Golden:2014xqa,Golden:2018gtk},  
namely 
\begin{align}
&w_3  =  x_{1}= \frac{\ab{23}\ab{14}}{\ab{12}\ab{34}} , \;\; 
 z_3  =  x_2= \frac{\ab{34} \ab{15}}{\ab{13}\ab{45}},  \;  \\ 
 & v_2 = \frac{1+x_3}{x_2 x_3}= \frac{ \ab{46}\ab{13}}{\ab{34}\ab{16}}\,, 
 \nn \\
 & 
 w_3 \xrightarrow{P} \bar{w}_3 =  \frac{\ab{14}\ab{56}}{\ab{45}\ab{16}} , \;\;   z_3 \xrightarrow{P} \bar z_3 =   \frac{\ab{16} \ab{24}}{\ab{46}\ab{12}} \,.
\end{align}
as well as their images under cyclic permutations. 

In addition we introduce
$r_1 =1/x_4= - \frac{\ab{14}{\ab{3I}}}{\ab{13}{\ab{4I}}}$ whose  images under $D_6$ transformations form a set of 12 cross ratios: 
\begin{align} 
\hskip-2.5mm 
r_a   = \frac{ \ab{a\,  a+3 } \ab{a+2\, I} }{ \ab{a \, a+2} \ab{I \, a+3 }},  \,
  \bar{r}_a  =    \frac{ \ab{a+5\,  a+2 } \ab{a+3\, I} }{ \ab{a+5 \, a+3} \ab{I \, a+2 }} , \hskip-2.5mm  
\end{align} 
satisfying $ 
 r_a \xrightarrow{P} r_{a+3},   r_a \xrightarrow{R} \bar r_a,  a+6 \doteq a$.

In terms of these ingredients, we can define $S \equiv \{r_1, w_1, z_1, v_1 , -w_1/\bar{w}_1,  -|z_1|^2\}$ plus their $D_6-$images and conclude that it has the desired properties, i.e. $\text{Li}_{1,2}(-S)$ (modulo products of logarithms) account for the one-loop symbol letters for the EEEC.

Next we investigate the structures of physical singularities, leading to a set of first-entry conditions that further constrain the function space
\begin{itemize}
\item Single-valueness in the physical domain away from the coplanar limit:  For $|\tau_1| =|\tau_2| =1,|s|< 1 \; ({\rm{or }}\; |s|>1)$, the function must be free from ambiguity in the principal values of  azimuthal angles, i.e. invariant as $\tau_1 \rightarrow e^{ \pm 2\pi i } \tau_1, \tau_2 \rightarrow e^{ \pm 2\pi i } \tau_2$. 
\item Near triple-collinear limit:  the function is free from logarithmic singularity in the triple-collinear limit as $s \rightarrow 0$ or $\infty$.   
\end{itemize}
As a consequence,  
 only 6 independent letters drawn from the set 
$ \left\{ \frac{\bar w_1}{ w_1} = \prod_{i=1}^3 \frac{1+w_i }{1+ \bar w_i } , \;  |z_i|^2 = \frac{v_i}{v_{i+1}}, \;  1+v_i  \right\}  $  can appear in the  the first entry.  In particular, a parity odd letter $\frac{\bar w_1}{ w_1}  = \frac{(s+\tau_1) (s+\tau_2) (1+s\tau_1 \tau_2) }{(1+s \tau_1)(1+s \tau_2)(s+\tau_1 \tau_2)} $ is allowed, which distiguishes the EEEC function space from the standard single-valued polylogarithms \cite{Dixon:2012yy}.   

In conclusion,  the one-loop EEEC function space comprises of classical polylogarithms whose arguments  drawn from the set $\{-S , 1+S\}$  satisfying the first-entry conditions. We observe that the final answer can be decompsed onto 14 such functions  as well as their cyclic permutations. 
Hence the one-loop EEEC in $\cN=4$ super Yang-Mills can be written in the a form that has manifest $D_6$ symmetry,
\begin{align} \label{Hmain}
\hskip-2.5mm
H^{\rm LO}_{\mathcal{N}=4} (\vec{n}_1, \vec n_2, \vec n_3 ) =  
 \sum_{i=1}^{14} b_{i} F_{i} +  \text{perm}(\vec n_1, \vec n_2, \vec n_3) 
\end{align}
where $\vec{b}$ is a set of rational functions of $(s, \tau_1, \tau_2)$, $\vec{F}$ contains weight-1 and weight-2 polylogarithmic functions. More explicitly, 
%
%
%
%
%
%
\begin{align} \label{Fset}
\vec{F} \equiv  \Big\{ f_1, f_2,f_3,  g_1, \cdots,  g_{11}  \Big\}  .
\end{align}
Each member of $\vec{F}$ has a distinct signature under the operation $\tau$ and $P$, where  $ \{ f_1, f_2, g_{2-4}, g_{8-11} \} $ and $ \{f_1, g_3,g_4,  g_7 , g_9 \}$ are odd under $\tau$ and $P$, respectively.   
The first three members $f_{1,2,3}$ are weight-1 functions, 
\begin{align}
f_1 =   \ln \frac{\bar w_1}{w_1}  , \quad  f_2 =  \ln |z_2|^2,\quad f_3=  \ln (1+ v_2)    
\end{align} 
The rest, $g_{1-11}$ are weight-2 functions, among which $g_{1-8}$ are characterized by the 9 $A_3-$cluster alphabets, depending only on  $ \{ w_i, z_i, v_i\}$ and their parity conjugation. 
\begin{align}
g_1  
&= {\rm{Li}}_2 (-  v_2)  \\
g_2  & = 
  \text{Li}_2 ( 1+  w_3 )  + \text{Li}_2  ( 1+  \bar w_3) +2\, {\rm{Li}}_2 (- v_3) \nn \\
  &  - 
\text{Li}_2  ( 1+ w_1  )  - \text{Li}_2   ( 1+ \bar w_1  ) -
 2\,{\rm{Li}}_2 ( - v_1  )   \nn \\
 g_3 &  =    \text{Li}_2( -z_2 ) -  \text{Li}_2(- \bar{z}_2 ) + \frac{1}{2} \ln   |z_2|^2 \ln \frac{1+ z_2}{ 1 + \bar z_2} \nn
\\
 g_{4} & =   \text{Li}_2 ( 1+ w_1) - \text{Li}_2  ( 1+\bar w_1 ) 
+ \text{Li}_2 ( 1+  w_2) \nn \\
& \;  - \text{Li}_2  (1+\bar w_2 )  +   \text{Li}_2 ( 1+ w_3 ) - \text{Li}_2 ( 1 + \bar  w_3)  \nn \\
 g_{5} &=  \pi^2\, \nn \\
  g_{6} &= \ln^2 \frac{ \bar{w}_1}{w_1} \nn \\ 
  g_{7} &=  \ln \frac{ \bar w_1}{w_1} \, \ln |z_2|^2     \nn \\
  g_{8} &=  \text{ln}\, (1+v_3)   \ln |z_1|^2
 -\text{ln}\,(1+v_1)  \ln |z_3|^2 \nn 
\end{align}
 $g_{11}$ is the only member depending on $\{r_i\}$, and the only one exhibiting an odd $Z_2-$signature: $g_{11} \xrightarrow{R} -g_{11}. $ 
\begin{align} 
 \hskip-0.2cm g_{11} & =  \sum_{i=1}^6 \text{Li}_2( -r_i ) -  \text{Li}_2(-\bar r_i ) + \frac{1}{2} \ln   | r_i |^2   \ln \frac{1+ r_i}{ 1 + \bar r_i} 
\end{align} 
The last two members $g_{9,10}$ are responsible for the homogeneous polynomials that appear as alphabet letters:
$ \ab{12}\ab{34}\ab{56}- \ab{23} \ab{45} \ab{61} , \ab{54}\ab{12}\ab{36}- \ab{41}\ab{23}\ab{65}  $. 
\begin{align} 
g_{9} & =  \frac12 \text{Li}_2 \Big(1-  \frac{\bar w_1}{w_1} \Big) - \frac12 \text{Li}_2 \Big(1-  \frac{ w_1}{\bar w_1} \Big)  \\
 g_{10}  &  =  \text{Li}_2 (1- |z_2|^2  ) + \frac12 \ln |z_2|^2 \ln |1-z_2|^2  
\end{align} 

In the ancillary files, we provide the explicit expressions for the coeffcients $\vec{b}$, as well as the full analytic expression for $H^{\rm LO}_{\mathcal{N}=4}$ in terms of the $\zeta_{ij}-$variables.  In FIG.~\ref{fig2} we display the function $H$ in various kinematic regions.

\medskip

\begin{figure}[t] 
\includegraphics[scale=0.5]{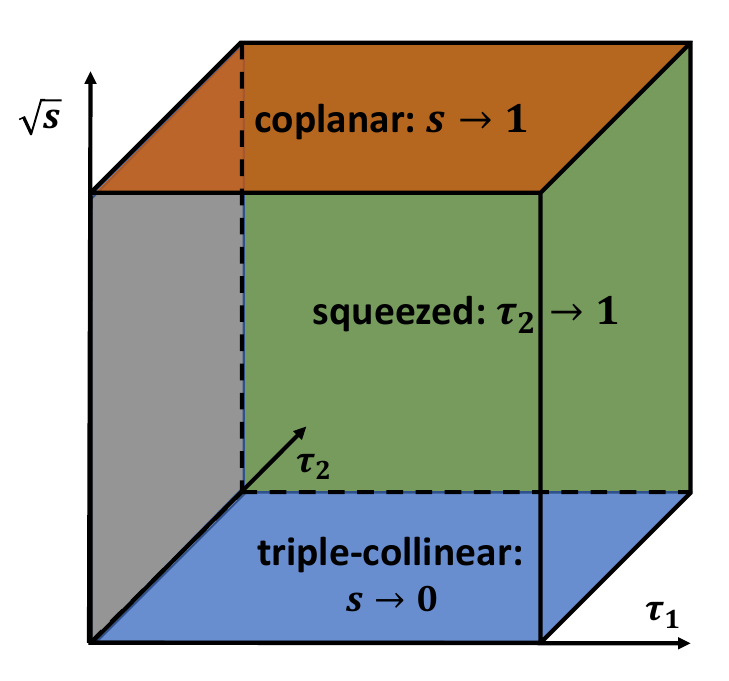}
\caption{
Kinematic regions for the EEEC and its singular regimes. }
 \label{fig:kimematic}
\end{figure}

\noindent \textbf{5.~ Special kinematics.}
 In the EEEC, physical singularities emerge on the surfaces in the three-dimensional parameter space displayed in FIG.~\ref{fig:kimematic}.
 Our rational parametrization Eq.~\eqref{anglecoords} makes it easy to access the three types of singular regions where the EEEC is enhanced, namely  the limit where three detectors are collinear ($ s \rightarrow $ or $\infty$), two of them are collinear  ($ \tau_2 \rightarrow 1 $) ,  or the three dectors are coplanar ($ s \rightarrow 1$).   
We extract analytically the leading power asymptotic behavious in all these limits.

\textbf{Triple-collinear limit} describes single jet events where the all three angles are small  $ \zeta_{ij} \sim 0$.   Taking $s \rightarrow 0$ ,  we verify that it is a regular limit free from logarithmic enhancement, such that $H  \sim \frac{1}{s^2} G( \tau_1, \tau_2)$.   Our expression for the function $G$  agrees with \cite{Chen:2019bpb}, uppon setting 
$ z=  \frac{1- \tau_1}{1-1/\tau_2},  \; \bar z =  \frac{1- 1/\tau_1}{1-\tau_2} $.

\textbf{Squeezed limit} correpsonds to the regime where we put two dectors on top of each other, $\zeta_{12} \sim 0 , \zeta_{13} \sim \zeta_{23} \sim \zeta$. 
We can access this limit by taking $\tau_2 \rightarrow 1$ keeping $s, \tau_1$ fixed,  and we find that the leading-power contributions can be grouped into a simple form
\begin{align} 
H \overset{\zeta_{12} \rightarrow 0}{\sim}  \frac{6}{\zeta_{12} } \left[ \frac{\zeta + \ln (1- \zeta)  }{(-1+ \zeta)\, \zeta^3}  \right] 
\end{align} 
where the coefficient of the leading pole depends on a single variable $\zeta= -\frac{ s(1-\tau_1)^2}{(1+s)^2 \tau_1}$. 

\textbf{Coplanar limit} correponds to the surface where $\Delta_1$ vanishes. 
In this regime, a soft particle recoils against three coplanar hard particles, which are scattered into two different half-planes.
We extract the leading singular behaviour by expanding the function $H$ around  $s =1$,  recalling $\tau_1= e^{i \phi_1}, \tau_2 = e^{i \phi_2} $,    
\begin{align}\label{Hcoplanar}
& \hskip-0.2cm H     \overset{ s \rightarrow 1 }{\sim }   18 \pi  \,
  \frac{  \theta\big( - \cos \frac{\phi_1}{2} \cos \frac{\phi_2}{2} \cos \frac{\phi_1 + \phi_2}{2} \big) }{ \big[ \sin \frac{\phi_1}{2} \sin \frac{\phi_2}{2} \sin \frac{\phi_1 + \phi_2}{2}  \big]^3 }   \times   \nn \\
&   \frac{1}{|1-s|}    \ln  \Big[    (1-s)^2  \tan^2 \frac{\phi_1}{2} \tan^2 \frac{\phi_2}{2} \tan^2 \frac{\phi_1+ \phi_2}{2}    \Big]   
\end{align}
The pole at $s=1$ comes solely from the discontinuity of  the $g_8$ function in the region where $  -w_{i}$ sit on the negative real axis.  
The physical origin  of the leading logarithms is  soft-collinear singularity.  
Lifting the result to $(4-2\epsilon)-$dimension, Eq.~\eqref{Hcoplanar} can be recast into distributional terms including $\delta(1-s)$ and plus-distributions \cite{Gelfand}.
 The $\epsilon-$IR divergences in the delta function cancels with those coming from the one-loop virtual contribution \cite{Bork:2011cj}, making the event shape finite at $s=1$. 
 
Since the $s \rightarrow 1$ behavior of the EEEC in $\mathcal{N}=4$ sYM is analogous to that in QCD, we anticipate that the leading Sudakov logarithms that appear in both theories are the same, which can be resummed to all loop orders \cite{Collins:1981uk}.

  \medskip
\noindent \textbf{6. Outlook.}  
Our work opens the way for several applications and further studies. 
Our one-loop formula Eq.~\eqref{Hmain} constitutes the first analytic result for a event-shape observable  living in a three-dimensional parameter space. 
Its symbol defines a set of 16 rational alphabet letters describing a finite physical observable.  
By embedding the kinematic space in a hexagon located on a unit circle, 
 we identify the symmetry properties and first-entry conditions that provide key information on the function space. 
Further studies on these mathematical structures and analytic properties will be crucial for  bootstrapping the observable at higher perturbative order in supersymmetric gauge theories or  in QCD. 

In addition to the leading asymptotic behaviours we provide, our result 
contains information about subleading powers as  well. 
The data in the  triple-collinear, squeezed and coplanar limit will shed new light on corresponding OPE limits of the light-ray operators \cite{Kravchuk:2018htv}, thus  making  it  possible to understand these limits at arbitrary coupling \cite{Chang:2022ryc,Chen:2022jhb}.  
In the meantime, the analysis in each aforementioned kinematic limts can be genralized to QCD, providing rich content for theoretical and phenomenological studies, as major process has been achieved in the triple-collinear limit \cite{Chen:2020adz,Chen:2021gdk,Chen:2020vvp,Komiske:2022enw}. 

Our approach to compute the EEEC in $\mathcal{N}=4$ sYM  benefits from the simplicity of the squared super form factor, which allows an integral representation  in a concise form.  
As novel research ideas emerged in recent studies on the form factors by means of harmonic superspace formalism \cite{Chicherin:2016fac,
Chicherin:2016fbj,FF2}, 
 modern amplitude techniques \cite{Lin:2021lqo,Guo:2021bym,
Dixon:2020bbt,
Dixon:2021tdw} and integrability descriptions \cite{Sever:2020jjx,
Sever:2021nsq,
Sever:2021xga},  
they open the way to probing the energy correlator observable at higher loop order or finite coupling,   
as well as relevant generalization of the event shape in quantum field theories \cite{Moult:2016cvt,Komiske:2017aww,Chivukula:2017nvl,Korchemsky:2021okt,
Martin:2020jlu} .

\medskip
\noindent \textbf{Acknowledgements.} We are grateful to Dmitri Chicherin, Gregory Korchemsky, Emery Sokatchev  and Alexander Zhiboedov for initiations and  generous help with the calculation.   
We acknowledge enlightening discussions with  Hao Chen,  Johannes Henn, Yibei Li,  Ian Moult, Alexander Tumanov, Tong-Zhi Yang  and Hua Xing Zhu.  
This  work is supported by  Shanghai Jiao Tong University (Grant No. WH220407213).  
In its early stage the work received funding from the European Research Council (ERC) under the European Union's Horizon 2020 research and innovation programme, {\it Novel structures in scattering amplitudes} (grant agreement No 725110).

\bibliographystyle{apsrev4-1} 

\bibliography{joh_more_refs}

\newpage

\onecolumngrid
\newpage
\appendix

\section*{Supplemental materials}

In the supplemental materials, we present the full analytic expression for three-point energy correlator EEEC in $\cN=4$ sYM.

\subsection{Analytic results for the $b_i$ coefficients}

The EEEC in $\cN=4$ sYM is completely described by the function space in Eq. (\ref{Hmain}), together with 14 rational coefficients. Recall that,
\begin{align}
    \text{ EEEC} (\chi_1, \chi_2, \chi_3) &\equiv  (8 \pi^2 )  \times  \Delta_1^{-1}\,  H(\vec{n}_1, \vec{n}_2 , \vec{n}_3 ) \notag\\
    &= (8 \pi^2 )  \times  \Delta_1^{-1}\left(\sum_{i=1}^{14}b_i F_i+  \text{perm}(\vec n_1, \vec n_2, \vec n_3) \right)
\end{align}
where there are two square roots in the expression, as shown in the letter:
\begin{align}
    \Delta_1&=\sqrt{\zeta_{12}^2+\zeta_{13}^2+\zeta_{23}^2-2\zeta_{12}\zeta_{13}-2\zeta_{12}\zeta_{23}-2\zeta_{13}\zeta_{23}+4\zeta_{12}\zeta_{13}\zeta_{23}}\notag \\
    \Delta_2&=\sqrt{\zeta_{12}^2+\zeta_{13}^2+\zeta_{23}^2-2\zeta_{12}\zeta_{13}-2\zeta_{12}\zeta_{23}-2\zeta_{13}\zeta_{23}}
\end{align}
The variables we introduce in section 4 can be written in terms of the standard $\zeta_{ij}-$variables, more explicitly,  
\begin{align}
w_1 &= \frac{-2 - \Delta_1 + \z_{12} + \z_{13} + \z_{23}}{2 (-1 + \z_{12}) (-1 + \z_{13})} , \quad z_1= \frac{-\Delta_1 + \z_{12} + \z_{13} - 2 \z_{12} \z_{13} -\z_{23}}{2 \z_{12} (-1 + \z_{13})},   \nn \\
v_1 &= \frac{\z_{23}}{1- \z_{23}}, \qquad r_1= -\frac{\Delta_2 - \z_{12} + \z_{13} - \z_{23}}{\Delta_1 + \Delta_2}\,.
\end{align}
as well as their images under cyclic shift: $i \rightarrow i+1,\, (i \doteq i+3) $,  Parity conjugation: $\Delta_1 \rightarrow -\Delta_1 $ and $R-$conjugation: $\Delta_2 \rightarrow -\Delta_2 $.

Finally the $b_i$ coefficients are given below.

\begin{align}
b_1 &= -\frac{ \Delta _1 \left(\zeta _{12}+\zeta _{13}+\zeta _{23}\right)}{3 \zeta _{12}^2 \zeta _{13}^2 \zeta _{23}^2} 
\quad 
b_2 = -\frac{2 \left(\zeta _{12}-\zeta _{23}\right)}{\zeta _{12} \zeta _{13}^2 \zeta _{23}}
\quad 
b_3 =-\frac{ \zeta _{12}^2-2 \zeta _{13} \zeta _{12}-2 \zeta _{23} \zeta _{12}+\zeta _{13}^2+\zeta _{23}^2-2 \zeta _{13} \zeta _{23} }{\zeta _{12}^2 \zeta _{13}^2 \zeta _{23}^2}
\end{align} 

\begin{align}
b_4 &= \frac{-2}{\left(\zeta _{12}-1\right) \zeta _{12}^2 \left(\zeta _{13}-1\right) \zeta _{13}^2 \left(\zeta _{12}+\zeta _{13}-1\right) \left(\zeta _{23}-1\right) \zeta _{23} \left(\zeta _{12}+\zeta _{23}-1\right) \left(\zeta _{13}+\zeta _{23}-1\right)} 
\left(-2 \zeta _{12} \zeta _{13}^6+2 \zeta _{12} \zeta _{23} \zeta _{13}^6  
\right.\nn \\
& \left. -\zeta _{23} \zeta _{13}^6+\zeta _{13}^6-2 \zeta _{12} \zeta _{23}^2 \zeta _{13}^5+2 \zeta _{23}^2 \zeta _{13}^5+5 \zeta _{12} \zeta _{13}^5-\zeta _{12}^2 \zeta _{23} \zeta _{13}^5-2 \zeta _{12} \zeta _{23} \zeta _{13}^5-2 \zeta _{13}^5-2 \zeta _{12} \zeta _{23}^3 \zeta _{13}^4-6 \zeta _{12}^2 \zeta _{23}^2 \zeta _{13}^4  \right.\nn \\
& \left.
+12 \zeta _{12} \zeta _{23}^2 \zeta _{13}^4-5 \zeta _{23}^2 \zeta _{13}^4-3 \zeta _{12} \zeta _{13}^4+10 \zeta _{12}^2 \zeta _{23} \zeta _{13}^4-11 \zeta _{12} \zeta _{23} \zeta _{13}^4+4 \zeta _{23} \zeta _{13}^4+\zeta _{13}^4+2 \zeta _{12} \zeta _{23}^4 \zeta _{13}^3-2 \zeta _{23}^4 \zeta _{13}^3\right.\nn \\
& \left.
-2 \zeta _{12}^2 \zeta _{23}^3 \zeta _{13}^3-2 \zeta _{12} \zeta _{23}^3 \zeta _{13}^3+5 \zeta _{23}^3 \zeta _{13}^3-3 \zeta _{12}^3 \zeta _{23}^2 \zeta _{13}^3+20 \zeta _{12}^2 \zeta _{23}^2 \zeta _{13}^3-12 \zeta _{12} \zeta _{23}^2 \zeta _{13}^3-19 \zeta _{12}^2 \zeta _{23} \zeta _{13}^3+16 \zeta _{12} \zeta _{23} \zeta _{13}^3
\right.\nn \\
& \left.
-3 \zeta _{23} \zeta _{13}^3+\zeta _{23}^5 \zeta _{13}^2-4 \zeta _{12} \zeta _{23}^4 \zeta _{13}^2+2 \zeta _{12}^2 \zeta _{23}^3 \zeta _{13}^2+9 \zeta _{12} \zeta _{23}^3 \zeta _{13}^2-4 \zeta _{23}^3 \zeta _{13}^2+6 \zeta _{12}^3 \zeta _{23}^2 \zeta _{13}^2-22 \zeta _{12}^2 \zeta _{23}^2 \zeta _{13}^2+3 \zeta _{23}^2 \zeta _{13}^2
\right.\nn \\
& \left.
+15 \zeta _{12}^2 \zeta _{23} \zeta _{13}^2-6 \zeta _{12} \zeta _{23} \zeta _{13}^2-\zeta _{23}^5 \zeta _{13}+2 \zeta _{12} \zeta _{23}^4 \zeta _{13}+2 \zeta _{23}^4 \zeta _{13}+4 \zeta _{12}^3 \zeta _{23}^3 \zeta _{13}-5 \zeta _{12}^2 \zeta _{23}^3 \zeta _{13}-4 \zeta _{12} \zeta _{23}^3 \zeta _{13}-\zeta _{23}^3 \zeta _{13}
\right.\nn \\
& \left.-10 \zeta _{12}^3 \zeta _{23}^2 \zeta _{13}+20 \zeta _{12}^2 \zeta _{23}^2 \zeta _{13}-9 \zeta _{12}^2 \zeta _{23} \zeta _{13}+2 \zeta _{12} \zeta _{23} \zeta _{13}-\zeta _{12}^2 \zeta _{23}^5+\zeta _{12} \zeta _{23}^5+4 \zeta _{12}^2 \zeta _{23}^4-4 \zeta _{12} \zeta _{23}^4+\zeta _{12}^4 \zeta _{23}^3-5 \zeta _{12}^3 \zeta _{23}^3
\right.\nn \\
& \left.
-2 \zeta _{12}^2 \zeta _{23}^3+5 \zeta _{12} \zeta _{23}^3+6 \zeta _{12}^3 \zeta _{23}^2-5 \zeta _{12}^2 \zeta _{23}^2-2 \zeta _{12} \zeta _{23}^2+2 \zeta _{12}^2 \zeta _{23}\right)  \nn \\
b_5 &= \frac{-\left(\zeta _{12}-\zeta _{23}\right)}{3 \left(\zeta _{12}-1\right) \zeta _{12}^3 \zeta _{13}^3 \left(\zeta _{12}+\zeta _{13}-1\right) \left(\zeta _{23}-1\right) \zeta _{23}^3 \left(\zeta _{13}+\zeta _{23}-1\right) } 
\left(\zeta _{12} \zeta _{13}^6+\zeta _{23} \zeta _{13}^6-\zeta _{13}^6-2 \zeta _{12}^2 \zeta _{13}^5  +2 \zeta _{12}^2 \zeta _{23}^2 \zeta _{13}^5
\right. \nn \\
& \left. +2 \zeta _{12} \zeta _{23}^2 \zeta _{13}^5-2 \zeta _{23}^2 \zeta _{13}^5+2 \zeta _{12}^2 \zeta _{23} \zeta _{13}^5-4 \zeta _{12} \zeta _{23} \zeta _{13}^5+2 \zeta _{13}^5+2 \zeta _{12} \zeta _{23}^3 \zeta _{13}^4+5 \zeta _{12}^2 \zeta _{13}^4+4 \zeta _{12}^2 \zeta _{23}^2 \zeta _{13}^4-11 \zeta _{12} \zeta _{23}^2 \zeta _{13}^4
\right. \nn \\
& \left. +5 \zeta _{23}^2 \zeta _{13}^4-4 \zeta _{12} \zeta _{13}^4+2 \zeta _{12}^3 \zeta _{23} \zeta _{13}^4-11 \zeta _{12}^2 \zeta _{23} \zeta _{13}^4+14 \zeta _{12} \zeta _{23} \zeta _{13}^4-4 \zeta _{23} \zeta _{13}^4-\zeta _{13}^4+2 \zeta _{12}^4 \zeta _{13}^3+2 \zeta _{12}^2 \zeta _{23}^4 \zeta _{13}^3-2 \zeta _{12} \zeta _{23}^4 \zeta _{13}^3
\right. \nn \\
& \left. +2 \zeta _{23}^4 \zeta _{13}^3-5 \zeta _{12}^3 \zeta _{13}^3-4 \zeta _{12}^3 \zeta _{23}^3 \zeta _{13}^3+2 \zeta _{12}^2 \zeta _{23}^3 \zeta _{13}^3+5 \zeta _{12} \zeta _{23}^3 \zeta _{13}^3-5 \zeta _{23}^3 \zeta _{13}^3+2 \zeta _{12}^4 \zeta _{23}^2 \zeta _{13}^3+2 \zeta _{12}^3 \zeta _{23}^2 \zeta _{13}^3-8 \zeta _{12}^2 \zeta _{23}^2 \zeta _{13}^3
\right. \nn \\
& \left. +4 \zeta _{12} \zeta _{23}^2 \zeta _{13}^3+3 \zeta _{12} \zeta _{13}^3-2 \zeta _{12}^4 \zeta _{23} \zeta _{13}^3+5 \zeta _{12}^3 \zeta _{23} \zeta _{13}^3+4 \zeta _{12}^2 \zeta _{23} \zeta _{13}^3-10 \zeta _{12} \zeta _{23} \zeta _{13}^3+3 \zeta _{23} \zeta _{13}^3-\zeta _{12}^5 \zeta _{13}^2+4 \zeta _{12}^2 \zeta _{23}^5 \zeta _{13}^2
\right. \nn \\
& \left. 
-2 \zeta _{12} \zeta _{23}^5 \zeta _{13}^2-\zeta _{23}^5 \zeta _{13}^2-12 \zeta _{12}^2 \zeta _{23}^4 \zeta _{13}^2+10 \zeta _{12} \zeta _{23}^4 \zeta _{13}^2+4 \zeta _{12}^3 \zeta _{13}^2-6 \zeta _{12}^3 \zeta _{23}^3 \zeta _{13}^2+23 \zeta _{12}^2 \zeta _{23}^3 \zeta _{13}^2-20 \zeta _{12} \zeta _{23}^3 \zeta _{13}^2+4 \zeta _{23}^3 \zeta _{13}^2
\right. \nn \\
& \left. 
-3 \zeta _{12}^2 \zeta _{13}^2+4 \zeta _{12}^5 \zeta _{23}^2 \zeta _{13}^2-12 \zeta _{12}^4 \zeta _{23}^2 \zeta _{13}^2+23 \zeta _{12}^3 \zeta _{23}^2 \zeta _{13}^2-25 \zeta _{12}^2 \zeta _{23}^2 \zeta _{13}^2+13 \zeta _{12} \zeta _{23}^2 \zeta _{13}^2-3 \zeta _{23}^2 \zeta _{13}^2-2 \zeta _{12}^5 \zeta _{23} \zeta _{13}^2\right. \nn \\
& \left. 
+10 \zeta _{12}^4 \zeta _{23} \zeta _{13}^2-20 \zeta _{12}^3 \zeta _{23} \zeta _{13}^2+13 \zeta _{12}^2 \zeta _{23} \zeta _{13}^2-\zeta _{12} \zeta _{23} \zeta _{13}^2+\zeta _{12}^5 \zeta _{13}+4 \zeta _{12}^3 \zeta _{23}^5 \zeta _{13}-4 \zeta _{12}^2 \zeta _{23}^5 \zeta _{13}-\zeta _{12} \zeta _{23}^5 \zeta _{13}+\zeta _{23}^5 \zeta _{13}
\right. \nn \\
& \left. 
-2 \zeta _{12}^4 \zeta _{13}-14 \zeta _{12}^3 \zeta _{23}^4 \zeta _{13}+16 \zeta _{12}^2 \zeta _{23}^4 \zeta _{13}-2 \zeta _{23}^4 \zeta _{13}+\zeta _{12}^3 \zeta _{13}+4 \zeta _{12}^5 \zeta _{23}^3 \zeta _{13}-14 \zeta _{12}^4 \zeta _{23}^3 \zeta _{13}+36 \zeta _{12}^3 \zeta _{23}^3 \zeta _{13}-32 \zeta _{12}^2 \zeta _{23}^3 \zeta _{13} \right. \nn \\
& \left. 
+5 \zeta _{12} \zeta _{23}^3 \zeta _{13}+\zeta _{23}^3 \zeta _{13}-4 \zeta _{12}^5 \zeta _{23}^2 \zeta _{13}+16 \zeta _{12}^4 \zeta _{23}^2 \zeta _{13}-32 \zeta _{12}^3 \zeta _{23}^2 \zeta _{13}+24 \zeta _{12}^2 \zeta _{23}^2 \zeta _{13}-4 \zeta _{12} \zeta _{23}^2 \zeta _{13}-\zeta _{12}^5 \zeta _{23} \zeta _{13}+5 \zeta _{12}^3 \zeta _{23} \zeta _{13}\right. \nn \\
& \left. 
-4 \zeta _{12}^2 \zeta _{23} \zeta _{13}+2 \zeta _{12}^3 \zeta _{23}^5-4 \zeta _{12}^2 \zeta _{23}^5+2 \zeta _{12} \zeta _{23}^5-4 \zeta _{12}^4 \zeta _{23}^4+4 \zeta _{12}^3 \zeta _{23}^4+4 \zeta _{12}^2 \zeta _{23}^4-4 \zeta _{12} \zeta _{23}^4+2 \zeta _{12}^5 \zeta _{23}^3+4 \zeta _{12}^4 \zeta _{23}^3-12 \zeta _{12}^3 \zeta _{23}^3
\right. \nn \\
& \left. 
+4 \zeta _{12}^2 \zeta _{23}^3+2 \zeta _{12} \zeta _{23}^3-4 \zeta _{12}^5 \zeta _{23}^2+4 \zeta _{12}^4 \zeta _{23}^2+4 \zeta _{12}^3 \zeta _{23}^2-4 \zeta _{12}^2 \zeta _{23}^2+2 \zeta _{12}^5 \zeta _{23}-4 \zeta _{12}^4 \zeta _{23}+2 \zeta _{12}^3 \zeta _{23}\right)  \nn \\
b_6 &=  \frac{2 \Delta _1}{ 3 \left(\zeta _{12}-1\right) \zeta _{12}^3 \left(\zeta _{13}-1\right) \zeta _{13}^3 \left(\zeta _{23}-1\right) \zeta _{23}^2 } 
\left(\zeta _{23} \zeta _{13}^4-\zeta _{13}^4+2 \zeta _{12} \zeta _{23}^2 \zeta _{13}^3-2 \zeta _{23}^2 \zeta _{13}^3-2 \zeta _{12} \zeta _{23} \zeta _{13}^3+\zeta _{23} \zeta _{13}^3+\zeta _{13}^3+\zeta _{23}^3 \zeta _{13}^2
\right. \nn \\
& \left. 
-2 \zeta _{12} \zeta _{23}^2 \zeta _{13}^2+\zeta _{23}^2 \zeta _{13}^2+2 \zeta _{12} \zeta _{23} \zeta _{13}^2-2 \zeta _{23} \zeta _{13}^2-\zeta _{23}^3 \zeta _{13}+\zeta _{12}^3 \zeta _{23}^2 \zeta _{13}-5 \zeta _{12}^2 \zeta _{23}^2 \zeta _{13}+3 \zeta _{12} \zeta _{23}^2 \zeta _{13}+\zeta _{23}^2 \zeta _{13}+4 \zeta _{12}^2 \zeta _{23} \zeta _{13}
\right. \nn \\
& \left. 
-3 \zeta _{12} \zeta _{23} \zeta _{13}+4 \zeta _{12}^2 \zeta _{23}^3-4 \zeta _{12} \zeta _{23}^3-4 \zeta _{12}^3 \zeta _{23}^2+4 \zeta _{12}^2 \zeta _{23}^2+4 \zeta _{12} \zeta _{23}^2-4 \zeta _{12}^2 \zeta _{23}\right) \nn \\
b_7 &=  \frac{-2 }{3 \Delta _1 \left(\zeta _{12}-1\right) \zeta _{12} \left(\zeta _{13}-1\right) \zeta _{13}^2 \left(\zeta _{23}-1\right) \zeta _{23}^3 } 
\left(\zeta _{13} \zeta _{12}^4-\zeta _{12}^4-4 \zeta _{13}^2 \zeta _{12}^3+3 \zeta _{13} \zeta _{12}^3+6 \zeta _{13}^2 \zeta _{23} \zeta _{12}^3 -8 \zeta _{13} \zeta _{23} \zeta _{12}^3
\right. \nn \\
& \left. 
+2 \zeta _{23} \zeta _{12}^3+\zeta _{12}^3+3 \zeta _{13}^3 \zeta _{12}^2-2 \zeta _{13}^2 \zeta _{12}^2+4 \zeta _{13}^3 \zeta _{23}^2 \zeta _{12}^2-12 \zeta _{13}^2 \zeta _{23}^2 \zeta _{12}^2+5 \zeta _{13} \zeta _{23}^2 \zeta _{12}^2-4 \zeta _{13} \zeta _{12}^2-6 \zeta _{13}^3 \zeta _{23} \zeta _{12}^2+8 \zeta _{13}^2 \zeta _{23} \zeta _{12}^2
\right. \nn \\
& \left. 
+5 \zeta _{13} \zeta _{23} \zeta _{12}^2-2 \zeta _{23} \zeta _{12}^2-\zeta _{13}^2 \zeta _{23}^3 \zeta _{12}+3 \zeta _{13}^2 \zeta _{12}+11 \zeta _{13}^2 \zeta _{23}^2 \zeta _{12}-8 \zeta _{13} \zeta _{23}^2 \zeta _{12}-8 \zeta _{13}^2 \zeta _{23} \zeta _{12}+3 \zeta _{13} \zeta _{23} \zeta _{12}+\zeta _{13} \zeta _{23}^2\right)
\nn \\
b_8 &= \frac{1}{  6 \left(\zeta _{12}-1\right) \left(\zeta _{13}-1\right) \zeta _{13} \left(\zeta _{12}+\zeta _{13}-1\right) \left(\zeta _{23}-1\right) \zeta _{23} \left(\zeta _{12}+\zeta _{23}-1\right) \left(\zeta _{13}+\zeta _{23}-1\right) } 
\left( -4 \zeta _{13}^2 \zeta _{12}^3+6 \zeta _{13} \zeta _{12}^3
\right. \nn \\
& \left. 
-2 \zeta _{13} \zeta _{23} \zeta _{12}^3-\zeta _{12}^3-4 \zeta _{13}^3 \zeta _{12}^2+26 \zeta _{13}^2 \zeta _{12}^2-24 \zeta _{13} \zeta _{12}^2-12 \zeta _{13}^2 \zeta _{23} \zeta _{12}^2+16 \zeta _{13} \zeta _{23} \zeta _{12}^2+3 \zeta _{12}^2-24 \zeta _{13}^2 \zeta _{12}-2 \zeta _{13}^2 \zeta _{23}^2 \zeta _{12}
\right. \nn \\
& \left. 
+32 \zeta _{13} \zeta _{12}+24 \zeta _{13}^2 \zeta _{23} \zeta _{12}-33 \zeta _{13} \zeta _{23} \zeta _{12}-3 \zeta _{12}-8 \zeta _{13}+9 \zeta _{13} \zeta _{23}+1  \right) \nn \\
b_9 &= \frac{1}{12 \left(\zeta _{12}-1\right) \zeta _{12} \left(\zeta _{13}-1\right) \zeta _{13}^2 \left(\zeta _{12}+\zeta _{13}-1\right) \left(\zeta _{23}-1\right) \zeta _{23}^2 \left(\zeta _{12}+\zeta _{23}-1\right) \left(\zeta _{13}+\zeta _{23}-1\right)}
\left( 4 \zeta _{13}^2 \zeta _{12}^5-4 \zeta _{13} \zeta _{12}^5
\right. \nn \\
& \left.
-8 \zeta _{13}^2 \zeta _{23} \zeta _{12}^5+8 \zeta _{13} \zeta _{23} \zeta _{12}^5-16 \zeta _{13}^2 \zeta _{12}^4+14 \zeta _{13} \zeta _{12}^4-8 \zeta _{13}^3 \zeta _{23} \zeta _{12}^4+36 \zeta _{13}^2 \zeta _{23} \zeta _{12}^4-30 \zeta _{13} \zeta _{23} \zeta _{12}^4+\zeta _{12}^4-4 \zeta _{13}^4 \zeta _{12}^3
\right. \nn \\
& \left.
+20 \zeta _{13}^3 \zeta _{12}^3+8 \zeta _{13}^2 \zeta _{12}^3+8 \zeta _{13}^3 \zeta _{23}^2 \zeta _{12}^3-16 \zeta _{13}^2 \zeta _{23}^2 \zeta _{12}^3-14 \zeta _{13} \zeta _{12}^3-4 \zeta _{13}^3 \zeta _{23} \zeta _{12}^3-24 \zeta _{13}^2 \zeta _{23} \zeta _{12}^3+36 \zeta _{13} \zeta _{23} \zeta _{12}^3-3 \zeta _{12}^3
\right. \nn \\
& \left.
-22 \zeta _{13}^3 \zeta _{12}^2+8 \zeta _{13}^3 \zeta _{23}^3 \zeta _{12}^2+12 \zeta _{13}^2 \zeta _{12}^2-56 \zeta _{13}^3 \zeta _{23}^2 \zeta _{12}^2+90 \zeta _{13}^2 \zeta _{23}^2 \zeta _{12}^2+6 \zeta _{13} \zeta _{12}^2+36 \zeta _{13}^3 \zeta _{23} \zeta _{12}^2-46 \zeta _{13}^2 \zeta _{23} \zeta _{12}^2
\right. \nn \\
& \left.
-21 \zeta _{13} \zeta _{23} \zeta _{12}^2+3 \zeta _{12}^2-30 \zeta _{13}^2 \zeta _{23}^2 \zeta _{12}-6 \zeta _{13} \zeta _{12}+4 \zeta _{13}^2 \zeta _{23} \zeta _{12}+23 \zeta _{13} \zeta _{23} \zeta _{12}-\zeta _{12}+2 \zeta _{13}-6 \zeta _{13} \zeta _{23} \right)\nn \\
b_{10}&= \frac{-\Delta _1 \left(\zeta _{23}-\zeta _{12}\right) }{6 \left(\zeta _{12}-1\right) \zeta _{12}^3 \zeta _{13}^2 \left(\zeta _{23}-1\right) \zeta _{23}^3 } 
\left(\zeta _{12}^3-2 \zeta _{13} \zeta _{12}^2+2 \zeta _{13} \zeta _{23} \zeta _{12}^2+\zeta _{23} \zeta _{12}^2-\zeta _{12}^2+\zeta _{13}^2 \zeta _{12}+2 \zeta _{13} \zeta _{23}^2 \zeta _{12}+\zeta _{23}^2 \zeta _{12}
\right. \nn \\
& \left.
+2 \zeta _{13} \zeta _{12}-4 \zeta _{13} \zeta _{23} \zeta _{12}-\zeta _{23} \zeta _{12}+\zeta _{23}^3-\zeta _{13}^2-2 \zeta _{13} \zeta _{23}^2-\zeta _{23}^2+\zeta _{13}^2 \zeta _{23}+2 \zeta _{13} \zeta _{23}\right) 
\nn \\
b_{11}&= \frac{-\left(\zeta _{12}-\zeta _{23}\right) }{ 2 \left(\zeta _{12}-1\right) \zeta _{12}^2 \left(\zeta _{13}-1\right) \zeta _{13}^3 \left(\zeta _{23}-1\right) \zeta _{23}^2 }
\left(2 \zeta _{13}^2 \zeta _{23} \zeta _{12}^3-2 \zeta _{13} \zeta _{23} \zeta _{12}^3-2 \zeta _{23} \zeta _{12}^3+2 \zeta _{12}^3-2 \zeta _{13}^2 \zeta _{23}^2 \zeta _{12}^2-4 \zeta _{13} \zeta _{23}^2 \zeta _{12}^2
\right. \nn \\
& \left.
+4 \zeta _{23}^2 \zeta _{12}^2-3 \zeta _{13} \zeta _{12}^2-\zeta _{13}^2 \zeta _{23} \zeta _{12}^2+10 \zeta _{13} \zeta _{23} \zeta _{12}^2-2 \zeta _{23} \zeta _{12}^2-2 \zeta _{12}^2+2 \zeta _{13}^2 \zeta _{23}^3 \zeta _{12}-2 \zeta _{13} \zeta _{23}^3 \zeta _{12}-2 \zeta _{23}^3 \zeta _{12}+\zeta _{13}^2 \zeta _{12}
\right. \nn \\
& \left.
-\zeta _{13}^2 \zeta _{23}^2 \zeta _{12}+10 \zeta _{13} \zeta _{23}^2 \zeta _{12}-2 \zeta _{23}^2 \zeta _{12}+3 \zeta _{13} \zeta _{12}-\zeta _{13}^2 \zeta _{23} \zeta _{12}-12 \zeta _{13} \zeta _{23} \zeta _{12}+4 \zeta _{23} \zeta _{12}+2 \zeta _{23}^3-\zeta _{13}^2-3 \zeta _{13} \zeta _{23}^2-2 \zeta _{23}^2
\right. \nn \\
& \left.
+\zeta _{13}^2 \zeta _{23}+3 \zeta _{13} \zeta _{23}\right)
\nn \\
b_{12} &= \frac{1}{ \Delta _1 \left(\zeta _{12}-1\right) \zeta _{12} \left(\zeta _{13}-1\right) \zeta _{13}^2 \left(\zeta _{23}-1\right) \zeta _{23}^2 }
\left( -\zeta _{12}^3+4 \zeta _{13} \zeta _{12}^2-4 \zeta _{13} \zeta _{23} \zeta _{12}^2+\zeta _{12}^2-2 \zeta _{13}^2 \zeta _{12}-4 \zeta _{13}^2 \zeta _{23}^2 \zeta _{12}-2 \zeta _{13} \zeta _{12}
\right. \nn \\
& \left.
+12 \zeta _{13}^2 \zeta _{23} \zeta _{12}-6 \zeta _{13} \zeta _{23} \zeta _{12}+2 \zeta _{13} \zeta _{23} \right) \nn \\
b_{13}&=  \frac{\left(\zeta _{23}-\zeta _{12}\right) }{\left(\zeta _{12}-1\right) \zeta _{12} \zeta _{13}^2 \left(\zeta _{23}-1\right) \zeta _{23} } 
\left(2 \zeta _{12}^2-2 \zeta _{23} \zeta _{12}-\zeta _{12}+2 \zeta _{23}^2-\zeta _{23}\right)  \nn \\
b_{14} &= \frac{1}{\Delta _2 \zeta _{13}^2 \zeta _{23}^3} \left(2 \zeta _{23} \zeta _{12}^2+2 \zeta _{12}^2-8 \zeta _{13} \zeta _{12}-8 \zeta _{13} \zeta _{23} \zeta _{12}-5 \zeta _{23} \zeta _{12}+6 \zeta _{13}^2+4 \zeta _{13}^2 \zeta _{23}+8 \zeta _{13} \zeta _{23}  \right) \,. 
\end{align}

In the ancillary file, we provide these coefficients and the transcendental function space in terms of the $\zeta_{ij}-$ variables. In particular, to meet the definition of logarithms' branch cuts in \texttt{Mathematica}, we slightly reorganize some of the polylogarithmic functions. As a check, we also set up the one-fold numerical integration in the file, which is in good agreement with our analytic result.

\end{document}